\documentclass[10pt,apj,twocolumn,twocolappendix,nofootinbib,notitlepage,numberedappendix,nobalancelastpage]{openjournal}

\usepackage{amsmath}
\usepackage{amssymb}
\usepackage[varg]{txfonts}
\usepackage{bm}
\usepackage{color}

\usepackage[colorlinks=true,
            urlcolor=blue,
            citecolor=blue,
            linkcolor=red,
            menucolor=blue,linktocpage=true]{hyperref}
\def\mrm#1{\mathrm{#1}}
\def\be{\begin{equation}}\def\ee{\end{equation}}
\def\bea#1\eea{\begin{align}#1\end{align}}
\def\x{{\mathbf{x}}}

\def\u{\mathbf{u}}

\def\k{{\mathbf{k}}}

\def\q{{\mathbf{q}}}

\def\v{{\mathbf{v}}}
\def\u{{\mathbf{u}}}

\def\dif{\mathrm{d}}

\newcommand\delD{\delta_\mathrm{D}}
\newcommand\im{\mrm{i}}
\def\ssp{\hspace{0.09em}}
\def\sp{\hspace{0.06em}}

\begin{document}

\title{Non-conservation and time non-locality of biased tracers\vspace{-15mm}}
\shorttitle{\sc{Non-conservation and time non-locality}}

\author{Lawrence Dam}
\affiliation{D\'epartement de Physique Th\'eorique,
University of Geneva, 24 Quai Ansermet, CH-1211 Geneva 4, Switzerland}

\begin{abstract}
\noindent 
We study the effect of ongoing formation and merger on the
assumed number conservation of biased tracers.
Using a Lagrangian approach, we present a model
of the number density which 
accounts for such effects. 
The model is nonlocal in time, reflecting the gradual assembly
of tracers from the underlying matter. 
The loss of tracers through merger is modelled by an
environmentally-dependent sink, such
that the merger rate is proportional to the local number density
(higher probability of an event in higher density regions).
We derive from our model a formula for the linear bias of non-conserved tracers,
showing that such tracers debias more rapidly than
conserved ones.
Over time the large-scale power becomes increasingly suppressed
relative to the conserved prediction,
behaviour which has been observed in simulations elsewhere.
Implications for current modelling approaches are discussed.
\end{abstract}


\maketitle

\section{Introduction}
\noindent 
The basic picture for the evolution of biased tracers, here called galaxies,
was described thirty years ago by \cite{Fry:1996fg}.
Supposing that the entire galaxy population formed
instantaneously at time $t_*$
and that galaxies then stream with the local matter flow,
Fry showed that the linear (Eulerian) bias $b_1$ tends to
approach unity over time:
\be\label{eq:b1E}
b_1(t;t_*)=1+[b_1(t_*)-1]\frac{D(t_*)}{D(t)}\,,
\ee
where $D(t)$ is the growth factor.
Independent of the initial
bias, galaxies evolve to trace the matter distribution;
high biases cannot be sustained.
\cite{Tegmark:1998} generalized this model to show that debiasing
is robust to ongoing galaxy formation and stochastic fluctuations at the
formation sites.
\cite{Chan:2012}, studying
the development of (space) nonlocal bias,
extended the modelling to higher order in perturbation theory.

This work provides a further expansion of the evolution model.
As with these later works, we also take into
account ongoing formation, though we will take a more overtly
Lagrangian approach.
The novel aspect of our model is that we also allow the loss of
tracers through a sink. The idea of this is that in regions
of high number density the probability of a merger event 
is higher. There are thus two competing effects in this model:
the addition of tracers through formation,
and their removal through merger.
%In the model we will present we retain the
%assumption that there is no velocity bias, since it
%is suppressed on large scales~\citep{bias_review}.

Our interest in bias evolution is motivated by
Lagrangian approaches to galaxy
clustering~\citep{Zeldovich:1969sb,Bernardeau:2001qr,Matsubara:2008wx,White:2014}.
In these approaches
there are 
two epochs of interest: the initial time $t_*$, when the Lagrangian bias
is laid down, and the final (or observed) time $t$. 
Assuming the number of galaxies is the same at both times, 
a continuity argument can be used to relate these two distributions,
namely by following each galaxy along its trajectory from the initial
to the final time.
This assumption is routine, but it is questionable~\citep{Espenshade:2024wtq}.
Formation and merger are ongoing processes 
that add and remove galaxies from the population. This causes
the mean comoving number density to evolve over time.
And unlike matter particles (whose total mass is conserved),
galaxies cannot be followed far into the past nor the future before
they lose meaning as distinct tracers. 
Concerning the Lagrangian approach the issue is that tracers observed
at time $t$ may
not yet exist at $t_*$, having only formed in the intervening period.
Conversely,
tracers that exist at $t_*$ may later merge with other objects,
leaving the sample that would be observed at time $t$.
These problems will be addressed in this work.

In Section~\ref{sec:noncon} we show how the Lagrangian approach
extends to non-conserved tracers, emphasizing the importance of
continuity.
Section~\ref{sec:model} presents a model of the number density
which takes into account the effects of
ongoing formation and merger. Corrections to the linear
bias~\eqref{eq:b1E} are computed and
implications for current modelling approaches are discussed.
Our conclusions can be found in Section~\ref{sec:conclusions}.

\section{Non-conservation: pushforward method}\label{sec:noncon}
\noindent
The Lagrangian approach to biased clustering begins with
the assumption that the total number of galaxies is conserved
between Lagrangian space $\q$ and Eulerian space $\x$.
Differentially, this is the statement
$n(\x,t)\sp\dif^3\x=n_{0}(\q)\sp\dif^3\q$, where $n_0$ is the
comoving number density at the initial time $t=0$.%
\footnote{In this work subscript `0' denotes the initial time.}
By following each galaxy $\q$ along its trajectory
$\x_t(\q)$ from its initial position $\x_0(\q)=\q$,
the (Eulerian) number density is
\be\label{eq:n}
n(\x,t)
=\int\dif^3\q\:\delD[\x-\x_t(\q)]\, n_{0}(\q)\,.
\ee
This formula, and its Fourier equivalent,
is often used in modelling the
power spectrum~\citep{Couchman_Bond88,Taylor:1996ne,1995MNRAS.273..475S,
Matsubara:2007wj,Chen:2020fxs},
the correlation function ~\citep{Bharadwaj:1996qm,Porciani:1996tn,White:2014,Vlah:2016bcl},
and methods of field-level inference~\citep{Schmittfull:2018yuk}.

The key to generalizing this pushforward approach
to non-conserved tracers
is to realize that Eq.~\eqref{eq:n} is an expression of
the continuity equation
\be\label{eq:cty}
\partial_t\ssp n + \frac1a\nabla\cdot(n\sp\v)=0\,,
\ee
where $a(t)$ is the scale factor and $\v(\x,t)$ is the galaxy
peculiar velocity.
In Appendix~\ref{app:cty-check} we show that
Eq.~\eqref{eq:n} formally solves the continuity equation.

Now the case of
non-conservation amounts to the addition of a source $S$
to the continuity equation~\eqref{eq:cty}. The source
describes the formation history and is generally
a complicated functional. Here we will treat it as an arbitrary
functional of the smoothed matter distribution.

\subsection{Non-conserved system} %Galaxy--matter co-evolution model
\noindent 
Consider a system in which the galaxy number density $n(\x,t)$ evolves
according to the continuity (or transport) equation
\be\label{eq:n-cty-S}
\partial_t\ssp n+\frac1a\nabla\cdot(n\sp\v_m)=S[\delta_m]\,,
\ee
for some arbitrary source $S[\delta_m]$.
Here $\delta_m(\x,t)$ is the matter overdensity and
$\v_m(\x,t)$ is the matter velocity field; these evolve
by the Euler--Poisson system~\citep{Bernardeau:2001qr}:
\bea\label{eq:lss-system}
\begin{split}
\partial_t\v_m+\frac1a\v_m\cdot\nabla\v_m+H\v_m=-\frac1a\nabla\phi\,, \\[-2pt]
\partial_t\ssp\delta_m+\frac1a\nabla\cdot[(1+\delta_m)\ssp\v_m]=0\,,
\quad\,
\nabla^2\phi=4\pi Ga^2\bar\rho_m\ssp\delta_m\,.
\end{split}
\eea
Here $\bar\rho_m(t)$ is the mean matter density,
$H(t)$ is the Hubble parameter,
$\phi(\x,t)$ is the gravitational potential, and
the total mass is conserved by virtue of the continuity equation.
These equations comprise a system of four equations for
$n,\delta_m,\v_m$, and $\phi$. Because
the Euler--Poisson system~\eqref{eq:lss-system}
is a closed subsystem, the dynamics of the galaxies
derive from the dynamics of the underlying matter field.
Co-evolution in this model means that galaxies locally
flow with matter, $\v=\v_m$. This is a good approximation
on scales large compared to the characteristic scales of
galaxy formation~\citep{Mirbabayi:2014zca,bias_review}.%
\footnote{We have skirted the issue that tracers
like haloes are composed of a collection of matter particles whose
internal motions are not coherent. In principle, one should
work with some weighted average.
However, from the perspective of effective field theory, if we give
up on a detailed description on small scales we may treat
tracers as if they were point particles,
identified for instance with the centre-of-mass
particle. These are select particles, or `marked points',
in the sea of all particles~\citep[e.g.][]{Sheth:2000ff}.
The advantage of this is that it allows us to identify the
motion of the tracer with the motion of these select particles.}
Note that while the dynamics of matter are nonlinear and self-consistent,
the dynamics of tracers are linear and passive in the sense
that once the Euler--Poisson system is solved, $\v_m$ is prescribed.

Note that when $S=0$ the above system of four equations recovers the 
galaxy--matter co-evolution model of \cite{Fry:1996fg}; see also
\cite{Baldauf:2012hs,Chan:2012,Mirbabayi:2014zca,Hui:2007zh}.

\subsection{Number density}
\noindent 
The solution to Eq.~\eqref{eq:n-cty-S} is%
\footnote{The general solution to Eq.~\eqref{eq:n-cty-S} also
contains the number-conserving homogeneous solution, in addition to
the particular solution~\eqref{eq:rho-nc}. 
We have neglected this since it corresponds to a pre-existing
population, which is unphysical.
}
\be\label{eq:rho-nc}
n(\x,t)
=\int^t_0\dif t_*\int\dif^3{\q}\,\delD[\x-\x_{t}(\q,t_*)]\, S(\q,t_*)\,,
\ee
where $\x_t(\q,t_*)$ is the trajectory of
tracer $\q$  initialized at time $t_*$.
In this model
$\dif N/\dif t={\dif}/{\dif t}\int\dif^3\x\, n(\x,t)
=\int\dif^3\q\, S(\q,t)\neq0$, so
number is not conserved.

Equation~\eqref{eq:rho-nc} generalizes the usual number-conserving
formula~\eqref{eq:n}, and
its form is in accordance with
% Duhamel's principle, 
the superposition principle for an inhomogeneous linear first-order ODE.
This says that
the particular solution~\eqref{eq:rho-nc} is built up
from the superposition of homogeneous solutions, each of which
solve the same initial value problem but with different initial conditions.
The idea is that the initial conditions can be
viewed as impulse sources at, for instance, time $t_*=0$ with
the subsequent $t>0$ evolution governed by the continuity
equation~\eqref{eq:cty}.
Thus by virtue of linearity the source can be viewed as a
collection of tightly-packed delta functions each localized
in time.

\subsection{Instantaneous formation}\label{sec:instant-formation}
\noindent
Galaxy formation is certainly an ongoing process, 
but it is instructive
to consider the limiting case in which
the entire population is formed instantaneously at, say,
$t_*=t_i$.
This amounts to inserting into Eq.~\eqref{eq:rho-nc}
the impulse source
\be\label{eq:S-instant}
S(\q,t_*)=\delD(t_*-t_i)\, n_i(\q)\,,
\ee
where $n_i(\q)\equiv n(\q,t_i)$ is the number density at the
time of formation. 
As is easy to check, this source recovers the familiar 
number-density formula~\eqref{eq:n}.

Note that this is also the source that recovers
the co-evolution model of \cite{Fry:1996fg,Mirbabayi:2014zca}
with initial data $n_i(\q)$.
Because of the delta function, the continuity equation 
$\partial_t\ssp n+\nabla\cdot(n\sp\v_m)/a=0$
is satisfied for all times, except $t=t_i$ when the
system undergoes a sharp kick, setting it in motion.
In other words, the tracers are suddenly brought into existence,
after which they can be treated as conserved particles,
advected by the matter flow.

To specify the model further the next step is to relate the
initial tracer density
to the underlying matter density, $n_i=n_i[\delta_m]$.
This is where the usual problem of galaxy {bias} enters,
here to be prescribed at the formation time (the Lagrangian bias).
The modern approach to
this problem is based on effective field theory,
whereby one expands $n_i[\delta_m]$ in terms of all operators
allowed by symmetry and the equivalence principle,
up to the desired perturbation order.

\subsection{Bias expansion and ongoing formation}\label{sec:expansion}
\noindent At this point it is interesting to compare the nonlocal-in-time,
nonperturbative formula~\eqref{eq:rho-nc}
with the nonlocal-in-time bias expansion of \cite{Senatore:2014eva,Angulo:2015eqa}.
In particular, here we will show this bias expansion is
contained in Eq.~\eqref{eq:rho-nc}.
First, since the expansion assumes the fluid approximation
we can invert the map $\q\mapsto\x=\x_t(\q,t_*)$.
Denote this inverse flow map
$\x\mapsto\q=\x_\mrm{fl}(\x,t;t_*)$ with
$\x_\mrm{fl}(\x,t;t)=\x$. Now we can do the integral over $\q$ in
Eq.~\eqref{eq:rho-nc} and write
\be
n(\x,t)=\int^t_0\dif t_* \,
	S\big(\x_\mrm{fl}(\x,t;t_*),t_*\big)\,.
\ee
(Here the Jacobian associated with the
inverse map has been absorbed into $S$.)
Assuming that $S$ respects the usual set of symmetries
of large-scale structure,
we may expand $S$ in terms of operators ${O}$
(e.g.\ the tidal field and the velocity gradient),
\be\nonumber
S\big(\x_\mrm{fl},t_*\big)
=\bar{n}(t_*)\sp H(t_*)\left(c\sp(t,t_*)+\sum_{O}c_O(t,t_*)\ssp O(\x_\mrm{fl},t_*)\right),
\ee
where the details of the formation history are now contained in
the response functions $c_O(t,t_*)$. 
With this expansion,
the tracer overdensity $\delta_g=(n-\bar{n})/\bar{n}$ 
with respect to the mean density
$\bar{n}(t)=\int\dif t_*\ssp \bar{n}(t_*)\ssp H(t_*)\ssp c(t,t_*)$ is
\be\label{eq:delta-nolocal-exp}
\delta_g(\x,t)=\sum_O\int^t_0\dif t_*\ssp H(t_*)\,
	c_O(t,t_*)\, O\big(\x_\mrm{fl}(\x,t;t_*),t_*\big)\,,
\ee
up to multiplicative functions of $t$ that
can always be absorbed into $c_O(t,t_*)$.
This is the nonlocal-in-time bias expansion. Often it is
expressed at the observation time $t$ by expanding the
operator about Eulerian position $\x$ and identifying the
bias coefficients as the time-averaged response functions.
However, it turns out that one has to carry the expansion to high
perturbative order to see the scale-dependent signatures of
time nonlocality.
Indeed, \cite{DAmico:2022ukl} showed up to fourth order
in perturbation theory,
after accounting for degeneracies between operators,
that the nonlocal-in-time expansion is
equivalent to the local-in-time expansion at the same order.%
\footnote{The local-in-time expansion is a limiting case of the
nonlocal-in-time expansion~\eqref{eq:delta-nolocal-exp}, and is
defined by local coefficients $c_O(t,t_*)=c_O(t)\,\delD(t-t_*)/H(t)$.}
More recently it was shown that nonlocal-in-time operators
appear at fifth~\citep{Donath:2023sav} and sixth order~\citep{Edison:2025bmj}.

\begin{figure}[t]
    \centering
    \includegraphics[width=1\columnwidth]{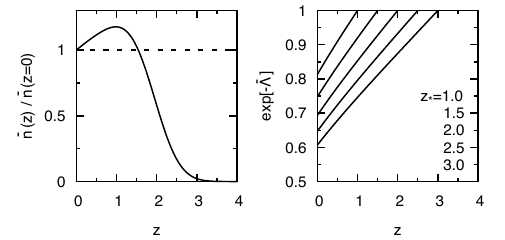}
    \caption{
      {\it Left}: evolution of the mean
      comoving density $\bar{n}(t)$
      normalized to the present value (here plotted against
      redshift for convenience).
      {\it Right}: evolution of the mean survival factor
      $\exp[-\bar\Lambda_t(t_*)]$
      for populations formed at $z_*=1$ to $z_*=3$ (from left to right).
      }
    \label{fig:functions}
\end{figure}

\section{Number density of non-conserved tracers}\label{sec:model}
\noindent We now present a model of the number density
which takes into account the effects
of ongoing formation and merger on clustering.
We then use it to estimate corrections to the
linear bias~\eqref{eq:b1E}.

\subsection{Model}
\noindent Consider the continuity equation [cf.~Eq.~\eqref{eq:n-cty-S}]
\be\label{eq:transport-model}
\partial_t\ssp n+\frac1a\nabla\cdot(n\sp\v_m)=J-\lambda\ssp n\,,
\ee
again coupled to the Euler--Poisson system~\eqref{eq:lss-system}.
Now the right-hand side has two contributions.
The first, $J(\x,t)\equiv S(\x,t)$, is as before a source function,
which adds tracers to the population.
This term is independent of $n$ and should be large in regions where
the matter density is high, these being the conditions needed for galaxy
formation to occur. The second contribution, $-\lambda(\x,t)\ssp n(\x,t)$, with
$\lambda>0$, is a sink term that models the loss of tracers, more strongly in regions where
the tracer number density is high, e.g.\ due to a higher probability of merger.
These two terms are in competition.
Initially, the number density is low and the first term dominates,
allowing the population to grow uninhibited. Population growth slows when
the number density becomes large enough that the second term
becomes important and tracers are lost through merger,
lowering the number density.

We will assume that $J=J[\delta_R]$ and $\lambda=\lambda[\delta_R]$, with
 $\delta_R$ the smoothed matter fluctuation. That is, we assume that formation and
merger are related in some (complicated) way to the underlying matter distribution.
In general, it will also depend on other variables of small-scale physics.
But provided $J$ and $\lambda$ are independent of the number density itself (no backreaction),
we do not need to know what these functions are; we can still write down
the formal solution to Eq.~\eqref{eq:transport-model}, since it remains a linear
differential equation in $n$ (treating $\delta_R$ and $\v_m$ as prescribed).
With the initial condition $n(\x,0)=0$ (that the population grows
from zero), the solution is
\bea\label{eq:n-gen}
&n(\x,t)=\int^t_0\dif t_*\!\ssp\int\dif^3\q\,\delD[\x-\x_t(\q,t_*)]\sp
	e^{-\Lambda_t(\q,t_*)}
	J(\q,t_*),
\eea
where $\Lambda_t(\q,t_*)\equiv\int^t_{t_*}\dif t'\lambda(\x_{t'}(\q,t_*),t')$.
Since $\Lambda_t>0$ the integrated or nonlocal factor $e^{-\Lambda_t(\q,t_*)}$
can be viewed as the survival `probability' that a tracer $\q$,
formed at $t_*$, persists until at least time $t$. In particular,
since $\Lambda_t$ depends on a particle's past history,
if the particle encounters more overdensities
it will have a lower survival probability
(higher chance of a merger event).

It is interesting to note that Eq.~\eqref{eq:n-gen}
closely resembles the equation
of radiative transfer for the intensity,
e.g.\ with $\Lambda_t(\q,t_*)$ analogous to the
optical depth~\citep{RybickiLightman}.

\begin{figure}[t!]
    \centering
    \includegraphics[width=1.0\columnwidth]{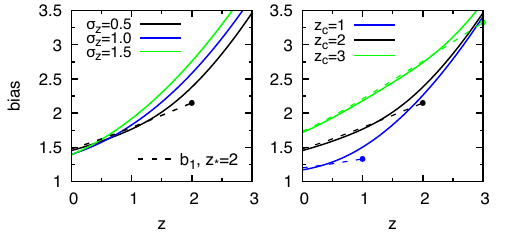}
    \caption{{\bf Source only.} Evolution of $b_{1,\mrm{eff}}$
    		[Eq.~\eqref{eq:beff-source-only}]
    		for different formation histories $\bar{J}$ as
			parametrized by $(\sigma_z,z_c)$.
    		{\it Left}: The impact on the bias of increasing the formation length $\sigma_z$,
			for fixed $z_c=2$ (solid lines). 
    		Also shown is the usual
			number-conserving bias~\eqref{eq:b1E}
		    that we would have if the tracer population was formed all
		    at once at $z_*=2$
		    (dashed lines).
			{\it Right}: The impact on the bias of changing the
			peak formation time $z_c$ for fixed formation length
			$\sigma_z=0.5$ (solid lines).
			For each curve we show the corresponding bias~\eqref{eq:b1E}
			of a conserved population
			(dashed lines).
		    }
    \label{fig:source-only}
\end{figure}

Equation~\eqref{eq:n-gen} is formally exact, but we will work
on large scales and expand $J[\delta_R]$ and $\lambda[\delta_R]$
to linear order in
$\delta_R$:
\begin{subequations}
\bea
J(\x,t)&=\bar{J}(t)\ssp\big[1+b_1(t)\ssp\delta_R(\x,t)\big]\,,\\
\lambda(\x,t)&=\bar{\lambda}(t)\ssp\big[1+c_1(t)\ssp\delta_R(\x,t)\big]\,.
\eea
\end{subequations}
Here we have four free functions:
$\bar J(t)$, the mean source rate;
$b_1(t)$, the linear bias;
$\bar\lambda(t)$, the mean loss rate;
and $c_1(t)$, the linear response of the sink to
a large-scale overdensity.
Note that allowing the source and sink to have environmental dependence
is crucial for there to be an effect on the
clustering and thus the bias, much as in peaks theory~\citep{BBKS}.
Note also that if we neglect the sink term
($\lambda=0$)
we recover the galaxy formation model of \cite{Tegmark:1998}.

The mean (comoving) density $\bar{n}$ is determined from these
functions and evolves as
$\dif\bar{n}/\dif t=\bar{J}(t)-\bar\lambda(t)\sp\bar{n}(t)$, 
i.e.\ Eq.~\eqref{eq:transport-model} at the background level.
Solving this equation using an integrating factor, we have
\be\label{eq:nbar}
\bar{n}(t)=\int^t_0\dif t_*\, e^{-\bar\Lambda_t(t_*)}\sp \bar{J}(t_*)\,,
\ee
where $\bar\Lambda_t(t_*)\equiv\int^t_{t_*}\dif t'\sp\bar\lambda(t')$.
For an initial population formed at once at $t_*=t_i$, so
$\bar{J}(t_*)=\bar{n}(t_i)\ssp\delD(t_*-t_i)$, this reduces to
$\bar{n}(t)=e^{-\bar\Lambda_t(t_i)}\sp\bar{n}(t_i)$.
In the usual case that number is conserved, this further reduces to
$\bar{n}(t)=\bar{n}(t_i)=\mrm{const}.$ for all times.

\subsection{Parametric model of source and sink}\label{sec:model-functions}
\noindent 
The source and sink functions are modelled (in redshift) as
\bea\label{eq:free}
\bar J(z)=J_0\ssp H(z)\ssp e^{-{(z-z_c)^2}/{2\sigma_z^2}}, 
\quad\;\;
\bar\lambda(z)=\lambda_0\ssp H(z)(1+z)^\gamma\,.
\eea
Here $z_c=2$ is the redshift of peak formation and $\sigma_z=0.5$
parametrizes the timescale of formation, with $\sigma_z\to0$ recovering
instantaneous formation. For the mean loss rate we set
$\lambda_0=0.25$ and $\gamma=0.5$.
We compute $b_1(t)$ for a population of haloes of mass $M=10^{12}M_\odot$
using the usual halo bias recipes~\citep{ShethTormen1999,ShethMoTormen2001}.
For simplicity we set $c_1(t)$ to unity since it is partially
degenerate with the scale parameter $\lambda_0$.
Note that $J_0$ does not enter into the
linear predictions of the overdensity below, and so will 
be left unspecified (it is however needed for the mean density~\eqref{eq:nbar}).
A \cite{Planck2018Cosmo} $\Lambda$CDM cosmology
is assumed for numerical work.

\subsection{Evolution of linear bias}
\noindent 
We now estimate the effect of non-conservation on the
linear bias $b_{1,\mrm{eff}}(t)$. Recall that this is
defined through
$\delta_g(\x,t)=b_{1,\mrm{eff}}(t)\ssp\delta_m(\x,t)$,
where $\delta_g=(n-\bar{n})/\bar{n}$
is the overdensity of the non-conserved tracer.
As we will see, non-conservation effects lead to corrections to the
debiasing relation~\eqref{eq:b1E},
which assumes a conserved tracer population instantaneously
brought into existence at $t=t_*$ with bias $b_1(t_*)$ at birth.
In all our comparisons below we will use a $t_*$ 
corresponding to $z_*=2$.

There are two competing effects, source vs sink, and it
will be instructive to study them in isolation before looking
at their combined effect.
Note that the formulae below are obtained by linearizing Eq.~\eqref{eq:n-gen}
about the mean density~\eqref{eq:nbar}.

{\it Source only}.
In the case of a nonzero source ${J}\neq0$ but no sink, $\lambda=0$,
we find%
\footnote{This expression was in essence also derived by \cite{Chan:2012}; see their 
equation 72 and note that in their notation
 $\dif n_*\leftrightarrow \bar{J}(t_*)\ssp\dif t_*$
and $\bar{n}(t)=\int^t_0\dif t_*\sp \bar{J}(t_*)$.}
\be\label{eq:beff-source-only}
b_{1,\mrm{eff}}(t)
=\frac{\int^t_0\dif t_*\sp \bar{J}(t_*)\ssp b_1(t,t_*)}
	{\int^t_0\dif t_*\sp \bar{J}(t_*)}\,,
\ee
where $b_{1}(t,t_*)$ is given by Eq.~\eqref{eq:b1E}.
Note that although we integrate down to $t=0$, the formation history 
is supported only on a narrow range set by $\bar{J}(t_*)$.

Figure~\ref{fig:source-only} shows the effective bias~\eqref{eq:beff-source-only}.
The left panel shows that the formation
of tracers over time, versus at a single time,
does not necessarily lead to a higher bias, as might be expected.
This will depend on the formation history.
While earlier-forming tracers are
initialized with higher bias, this may not be enough to
overcome the effect of later-forming ones, whose bias,
though initially lower, decreases more slowly~\citep{Fry:1996fg}.
As shown in the left panel in Figure~\ref{fig:source-only},
these effects largely average out in the model by the time
we reach $z\simeq0$, i.e.\ after the epoch of formation has ended.
Separately, the right panel shows that changing the
peak formation redshift $z_c$ (keeping $\sigma_z=0.5$ fixed)
leads only to a small difference compared to the usual prediction~\eqref{eq:b1E}.

\begin{figure}[t!]
    \centering
    \includegraphics[width=1\columnwidth]{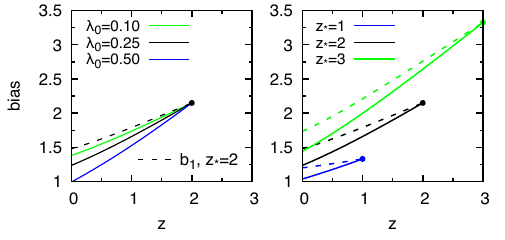}
    \caption{{\bf Sink only.} 
    		Evolution of $b_{1,\mrm{eff}}$ [Eq.~\eqref{eq:b1E-sink}]
			for different parameter values.
			 {\it Left}: The impact of the present-day mean loss parameter
			 $\lambda_0$ on $b_{1,\mrm{eff}}$ (solid lines).
			 The usual number-conserving bias~\eqref{eq:b1E} is
			 also shown (dashed lines).
		 	 {\it Right}: The impact of the formation time $z_*$ on
			 $b_{1,\mrm{eff}}$ (solid lines). 
		     For each $z_*$, indicated by filled circles, 
		     there are two evolutionary tracks the bias can take,
		     depending on whether the tracer population is conserved (dashed lines)
		     or not (solid lines).}
    \label{fig:loss}
\end{figure}

{\it Sink only}.
In the case of a nonzero sink $\lambda\neq0$ and no ongoing formation,
$\bar{J}(t_*)=\bar{n}(t_i)\ssp\delD(t_*-t_i)$, we find
\be\label{eq:b1E-sink}
b_{1,\mrm{eff}}(t)
=b_1(t,t_i)+\Delta b_\lambda(t,t_i)\,,
\ee
where $b_{1}(t,t_i)$ is given by Eq.~\eqref{eq:b1E} and
\be\label{eq:Delta-b}
\Delta b_\lambda(t,t_i)
\equiv
	-\int^t_{t_i}\dif t'\ssp\bar\lambda(t')\ssp c_1(t')\ssp\frac{D(t')}{D(t)}
\ee
is a negative correction.
The bias~\eqref{eq:b1E-sink} is shown in Figure~\ref{fig:loss}.
In our fiducial model (shown in solid black) $\Delta b_\lambda$ yields
a correction of about $20\%$ compared to the usual $b_1$.
The right panel shows that these corrections are fairly robust to
different formation times $z_*$.
Unlike the source-only bias~\eqref{eq:beff-source-only}, 
it is much easier to obtain a sizeable correction with the sink.

{\it Source and sink}.
The combined formula is
\bea
&{b}_{1,\mrm{eff}}(t)
=\frac{\int^t_0\dif t_*\ssp e^{-\bar{\Lambda}_t(t_*)}\bar{J}(t_*)\,
	\big[b_1(t,t_*)+\Delta b_\lambda(t,t_*)\big]}
	{\int^t_0\dif t_*\, e^{-\bar{\Lambda}_t(t_*)}\bar{J}(t_*)}\,, \label{eq:beff}
\eea
where $\Delta b_\lambda$ is given by Eq.~\eqref{eq:Delta-b} and
$\bar{\Lambda}_t(t_*)$ is given by Eq.~\eqref{eq:nbar}.
The biases~\eqref{eq:beff-source-only} and \eqref{eq:b1E-sink}
are special cases of Eq.~\eqref{eq:beff}:
the former is recovered when $\bar\lambda=0$ and $c_1=0$
(so $\bar\Lambda=0$ and $\Delta b_\lambda=0$), while
the latter when $\bar{J}(t_*)=\bar{n}(t_i)\ssp\delD(t_*-t_i)$.

The effective bias~\eqref{eq:beff} can be read as follows.
Ongoing formation continually injects new cohorts of tracers with
initial bias $b_1(t_*)$,
while gravitational evolution debiases older cohorts in the usual
$1/D(t)$ way.
Late forming cohorts have had less time to debias,
and so sustained late-time formation tends to keep the effective bias
higher than it would otherwise be had the population formed entirely at early times.
Furthermore, the survival factor $e^{-\bar\Lambda}$ further suppresses
the contribution
of older cohorts, thereby upweighting newly formed cohorts in
Eq.~\eqref{eq:beff}, which begin life with $e^{-\bar\Lambda}=1$.
In addition, if the removal rate of tracers $\lambda$ is correlated
with the matter density field, so $c_1\neq0$, then each cohort receives
an additional correction $\Delta b_\lambda$. If removal is more efficient
in overdense regions then $c_1>0$ gives $\Delta b_\lambda<0$, thus lowering
the bias $b_1$ expected if the cohort was conserved.
Preferential removal in overdense
environments largely outweighs the enhanced formation bias of tracers
that were formed in those same environments.

\begin{figure}[t!]
    \centering
    \includegraphics[width=0.85\columnwidth]{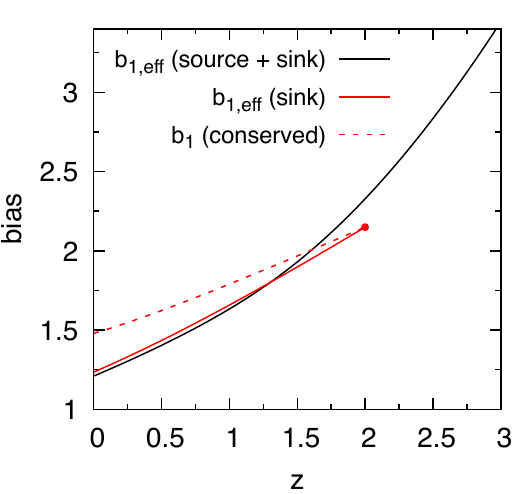}
    \caption{{\bf Source and sink.} Evolution of
    		 $b_{1,\mrm{eff}}$ [Eq.~\eqref{eq:beff}] 
		 including both source and sink (solid black).
		     The sink-only bias (solid red) shows that the decrement
		     is largely explained by
		     population loss (through $\Delta b_\lambda$) and that
				ongoing formation has a small effect.
			    	The number-conserving bias~\eqref{eq:b1E}
		     is shown for reference (dashed red).
}
    \label{fig:beff}
\end{figure}

Figure~\ref{fig:beff} shows the effective bias for our parametric model.
Here it is seen that even after taking into account
the formation history $\Delta b_\lambda$ remains
the largest correction, with small difference between the sink-only 
bias~\eqref{eq:b1E-sink} and the total bias~\eqref{eq:beff}.

The environmental dependence of $\lambda$ gives rise to
$\Delta b_\lambda$ and hence the large correction.
(Our first and failed attempt at a model of $\lambda$
allowed time dependence only.)
Without the scale dependence, through $\delta_R$,
the survival factor
$e^{-\Lambda}$ modifies the density, but not
the density \emph{fluctuation} nor the bias~\eqref{eq:b1E-sink}.
That the bias is unaffected is essentially for the same reason
that in peaks theory~\citep{BBKS,Kaiser:1984sw}
enhanced clustering is seen for a {select} population of tracers
(e.g.\ above some density threshold), but not for the total
population.

\subsection{Modelling implications}
\noindent
The validity of number conservation has also been investigated
by \cite{Espenshade:2024wtq}. Using $N$-body simulations,
they showed that the pushforward formula 
[Eq.~\eqref{eq:deltag} below]
overestimates by a factor of three
the large-scale power spectrum of haloes in the mass bin
$11.82<\log (M/h^{-1}M_\odot)<12.32$.
This corresponds to a $b_1(z=0)$ which is
about $60\%$ that of the number-conserving prediction~\eqref{eq:b1E}.

This suppression is explained by population loss 
through halo merger. In the model we have presented the mechanism
by which this occurs is through the sink. This removes tracers
from the population such that the population debiases more
rapidly.
As Figure~\ref{fig:beff} shows, the overall bias
is generally lower than the conserved prediction.

We note however that our model
underestimates the amount of suppression: we find that
$b_{1,\mrm{eff}}$ is about $80\%$ the value of $b_1$, i.e.\
above the $60\%$ target needed for a factor-of-three
suppression in power.
Besides centring the formation history at $z_c=2.9$
and using the appropriate mass bin,
we have not however attempted a more realistic model.
Certainly, the suppression can be brought down to the required level
by tweaking the free functions $\bar\lambda(t)$ and $c_1(t)$.
Whether this would yield a realistic merger history is another
matter. We leave this for future work.

\cite{Espenshade:2024wtq} also raised concerns about the
validity of certain formulae used in modelling the power spectrum
of tracers, in particular
\be\label{eq:deltag}
\delta_g(\k,t)
= \int\dif^3\q\,
  {e}^{-\im\k\cdot(\q+\bm\psi(\q,t))} [1+\delta_{g0}(\q)]\,,
\ee
where $\bm\psi(\q,t)=\x_t(\q)-\q$ is the displacement field
and $\delta_{g0}(\q)$ is the initial overdensity.
This expression follows from
Eq.~\eqref{eq:n}, which assumes number conservation.

Provided one treats Eq.~\eqref{eq:deltag} effectively,
this formula remains valid for non-conserved tracers.
To see this let us
obtain the analogous formula, accounting for population evolution.
For simplicity, let us focus on the effect of the
sink and ignore the source (since we have already discussed it
in Section~\ref{sec:expansion}).
Thus putting $J(\q,t_*)=n_0(\q)\ssp\delD(t_*)$ in
Eq.~\eqref{eq:n-gen}, writing
$n=\bar{n}(1+\delta_g)$ where $\bar{n}(t)=e^{-\bar\Lambda(t)}\ssp \bar{n}_0$
is the mean density, and then taking the Fourier transform, 
we have
\be
\delta_g(\k,t)
= \int\dif^3\q\,
  {e}^{-\im\k\cdot(\q+\bm\psi(\q,t))} Q(\q,t)\ssp [1+\delta_{g0}(\q)]\,,
\ee
where $Q(\q,t)\equiv e^{-\Lambda_t(\q)+\bar\Lambda(t)}$.
Notice that the only difference from Eq.~\eqref{eq:deltag} is
the multiplicative factor $Q$; the functional form is unchanged.

One way to deal with $Q$ 
is to associate it with the plane wave
$e^{-\im\k\cdot(\q+\bm\psi)}$. Since both factors depend
on the past history of the tracer, both are time nonlocal quantities
related to propagation.
However, to preserve
the usual modelling approach, based on operator 
expansions~\citep{Vlah:2016bcl,Schmittfull:2018yuk,Chen:2020fxs},
a better way is to associate $Q$ with $1+\delta_{g0}$.
This is because both $Q$ and $\delta_{g0}$
are uncertain functions that depend in some complicated way on
the underlying matter field $\delta_m$.
Provided $Q$ respects the same symmetries as $\delta_{g0}$,
we can treat the `dressed density' $F[\delta_m]\equiv Q(\q,t)[1+\delta_{g0}(\q)]$
as a single function, and expand it in the usual
operators~\citep{bias_review}.
Simply, $Q$ has the effect of renormalizing the
bias coefficients.

\section{Conclusions}\label{sec:conclusions}
\noindent 
Pushforward methods are not limited to conserved tracers;
they also extend to non-conserved tracers.
The key is the continuity equation.
By solving it with (arbitrary) environmentally-dependent
source and sink functions,
we obtained an expression for the number density~\eqref{eq:n-gen}
that evolves over time due to the continual addition of
tracers produced from the underlying matter,
and the loss of tracers through merger events in regions of high
number density.
The usual number-conserving density~\eqref{eq:n} is recovered
from the general formula~\eqref{eq:n-gen} in the limit of
an impulse source (instantaneous formation) and vanishing sink.
For an arbitrary source function, expressed in terms of operators,
Eq.~\eqref{eq:n-gen} recovers the nonlocal-in-time bias expansion.

We studied the impact of the formation history and merger on the evolution of
linear bias. The largest correction $\Delta b_\lambda$ is
related to the sink's environmental dependence, without which we found
there would be little effect on the bias evolution.
This correction leads to faster debiasing than the
standard prediction~\eqref{eq:b1E}, so that
by $z=0$ the bias, for our particular model, is about $20\%$ lower than
if that same population was conserved.
In other words, non-conserved tracers are less biased tracers
than conserved ones.
This means that over time the large-scale power is increasingly
suppressed with respect to the evolution of
power for a conserved tracer. We caveat however that these 
estimates depend on the formation history, for which
we implemented a simple yet plausible parametric model.
No attempt was made at a more realistic model.

\begin{acknowledgements}
LD thanks Anton Chudaykin and Omar Darwish for useful discussions.
This work was partially supported by the European Research Council under
the European Union’s Horizon 2020 Research and Innovation Programme
(grant agreement no.~863929).
\end{acknowledgements}

%~~~~~~~~~~~~~~~~~~~~~~~~~~~~~~~~~~~~~~~~~~~~~~~~~~~~~~~~~~~~~
\bibliography{main}
%~~~~~~~~~~~~~~~~~~~~~~~~~~~~~~~~~~~~~~~~~~~~~~~~~~~~~~~~~~~~~

\appendix

\section{Solution to the continuity equation}\label{app:cty-check}
\noindent To see that the density~\eqref{eq:n}
formally solves the continuity equation,
differentiate Eq.~\eqref{eq:n} with respect to $t$:
\be\label{eq:dn}
\partial_t\ssp n(\x,t)
=-\int\dif^3\q\frac{\dif\x_t(\q)}{\dif t}\cdot\nabla_\x\ssp
\delD[\x-\x_t(\q)]\sp n_{0}(\q)\,.
\ee
Here $\dif\x_t/\dif t$ is the velocity of particle $\q$.
Now the Eulerian velocity is given by
the mean particle velocity at $\x$:
\be\label{eq:v}
\u(\x,t)
\equiv \frac{1}{n(\x,t)}{\int\dif^3\q\, a(t)\, \dfrac{\dif\x_t(\q)}{\dif t}\,
    \delD[\x-\x_t(\q)]\ssp n_{0}(\q)}\,,
\ee
where $n(\x,t)$ is given by Eq.~\eqref{eq:n},
and the appearance of $a(t)$ is because $\u$ is the physical velocity.
By working Eq.~\eqref{eq:v} into Eq.~\eqref{eq:dn},
taking the gradient outside the integral, and bringing all terms
to one side, we have
\be\label{eq:cty-u}
\partial_t\ssp n+\frac1a\nabla_\x\cdot(n\sp\u)=0\,,
\ee
which is the continuity equation.
Note that Eq.~\eqref{eq:n} requires that the particle ensemble
begins in a cold state (negligible velocity dispersion).
However, single streaming need not be assumed in the subsequent
evolution of the system. 

If we assume single streaming, as in a fluid description,
tracers will flow according to the velocity field $\v_m(\x,t)$,
\bea\label{eq:eom-ss}
\frac{\dif\x_t(\q)}{\dif t}
=\frac1a\,\v_m(\x_t(\q),t)\,,
\eea
where each Eulerian position $\x=\x_t(\q)$ corresponds to
a unique $\q$. Inserting this into Eq.~\eqref{eq:v}
yields $\u(\x,t)=\v_m(\x,t)$, and Eq.~\eqref{eq:cty-u} can then
be read as the continuity equation of a fluid.

\end{document}